\documentclass[usenatbib]{mn2e}
\usepackage[letterpaper,margin=0.8in]{geometry}
\usepackage{hyperref}
\usepackage{natbibmnfix,fixltx2e}
\usepackage{astrojournals}
\usepackage{mathtools}
\usepackage{amsmath}
\usepackage{txfonts}
\usepackage{graphicx}
\usepackage{microtype}
\usepackage{xspace}
\usepackage{xcolor}
\usepackage[utf8]{inputenc}

\hypersetup{%
  pdftitle={title},
  pdfauthor={\textcopyright\ authors},
  bookmarksopen=true,
  colorlinks=true,
  linkcolor=black,
  citecolor=black,
  urlcolor=black}

\date{}
\pagerange{\pageref{firstpage}--\pageref{lastpage}}
\pubyear{2015}

\renewcommand*{\vec}[1]{\boldsymbol{#1}}
\newcommand{\grad}{\vec{\nabla}}
\renewcommand{\dot}{\vec{\cdot}}

\begin{document}
\label{firstpage}

\title[Kelvin-Helmholtz Benchmark]{A Validated Nonlinear Kelvin-Helmholtz Benchmark for Numerical Hydrodynamics}

\author[Lecoanet et al]{D. Lecoanet$^{1,2}$\thanks{E-mail: dlecoanet@berkeley.edu}, M. McCourt$^{3}$, E. Quataert$^{1,2}$, K. J. Burns$^{4}$,
\newauthor G. M. Vasil$^{5}$, J. S. Oishi$^{6}$, B. P. Brown$^{7}$, J. M. Stone$^{8}$, \& R. M. O'Leary$^{9}$ \\
$^{1}$Physics Department, University of California, Berkeley, CA 94720, USA\\
$^{2}$Astronomy Department and Theoretical Astrophysics Center, University of California, Berkeley, CA 94720, USA\\
$^{3}$Institute for Theory and Computation, Center for Astrophysics, Harvard University, 60 Garden St, Cambridge, MA 02138, USA\\
$^{4}$Department of Physics, Massachusetts Institute of Technology, Cambridge, Massachusetts 02139, USA\\
$^{5}$School of Mathematics \& Statistics, University of Sydney, NSW 2006, Australia\\
$^{6}$Department of Physics, Farmingdale State College, Farmingdale, NY 11735, USA\\
$^{7}$Laboratory for Atmospheric and Space Physics and Department of Astrophysical \& Planetary Sciences, University of Colorado, Boulder, Colorado 80309, USA\\
$^{8}$Department of Astrophysical Sciences, Princeton University, Princeton, NJ 08544, USA\\
$^{9}$JILA, University of Colorado and NIST, 440 UCB, Boulder, CO 80309-0440, USA\\
}

\maketitle

\begin{abstract}
The nonlinear evolution of the Kelvin-Helmholtz instability is a popular test for code verification. To date, most Kelvin-Helmholtz problems discussed in the literature are ill-posed: they do not converge to any single solution with increasing resolution.  This precludes comparisons among different codes and severely limits the utility of the Kelvin-Helmholtz instability as a test problem. The lack of a reference solution has led various authors to assert the accuracy of their simulations based on ad-hoc proxies, e.g., the existence of small-scale structures.   This paper proposes well-posed Kelvin-Helmholtz problems with smooth initial conditions and explicit diffusion.  We show that in many cases numerical errors/noise can seed spurious small-scale structure in Kelvin-Helmholtz problems.  We demonstrate convergence to a reference solution using both Athena, a Godunov code, and Dedalus, a pseudo-spectral code. Problems with constant initial density throughout the domain are relatively straightforward for both codes.  However, problems with an initial density jump (which are the norm in astrophysical systems) exhibit rich behavior and are more computationally challenging.   In the latter case, Athena simulations are prone to an instability of the inner rolled-up vortex; this instability is seeded by grid-scale errors introduced by the algorithm, and disappears as resolution increases.  Both Athena and Dedalus exhibit late-time chaos.    Inviscid simulations are riddled with extremely vigorous secondary instabilities which induce {\it more} mixing than simulations with explicit diffusion. Our results highlight the importance of running well-posed test problems with demonstrated convergence to a reference solution. To facilitate future comparisons, we include the resolved, converged solutions to the Kelvin-Helmholtz problems in this paper in machine-readable form.
\end{abstract}


\section{Introduction}
\label{intro}

The Kelvin-Helmholtz (KH) instability results from a wide array of velocity-shear profiles 
in a continuous fluid, or across the interface between two
distinct fluids.  The instability is ubiquitous in nature, playing
important roles in meteorology, oceanography, and engineering.
The KH instability plays a particularly prominent role in
astrophysical systems ranging in scale
from stellar interiors \citep[e.\,g.][]{Bruggen2001} and
protoplanetary disks \citep[e.\,g.][]{Johansen2006} to the evolution
of the intergalactic medium \citep[e.\,g.][]{Nulsen1982,Nulsen1986}.
Physically, the KH instability wraps up coherent sheets
of vorticity into smaller, less organized structures. The small scale motion then
stretches and cascades to yet smaller scales. The
instability therefore plays fundamental roles in fluid mixing and in the
transition to turbulence.

Because of its prevalence in nature and its physical significance, KH test problems are commonly used to evaluate the accuracy of different astrophysical hydrodynamics codes
\citep[e.\,g.][]{Springel2010,Hopkins2015,Schaal2015}: if a code can
properly simulate the KH instability, it is presumed to
capture mixing and turbulence in astrophysical simulations.
Ideally, such an important test problem should stand against an
analytic solution to ensure the veracity (not just reproducibility) of
simulation results.  Some analytic work addresses the KH instability with a sheet vortex model
\citep{Moore1979}, but only for incompressible fluid equations.  For the compressible Navier-Stokes equations relevant to
astrophysics, no analytic description of the nonlinear
KH instability currently exists.

Absent a nonlinear analytic prediction, a resolved reference simulation provides the only reasonable approximation of the true solution. Comparing to a well-controlled and high-resolution benchmark gives a proxy for the true error of a given test.
\citet{Robertson2010} and \citet{McNally2012} present careful studies
of the early evolution of the KH instability. These
authors also point out the numeric ill-posededness of contact-discontinuity simulations, in spite of existing analytical solutions in the linear and/or incompressible regimes.  These works emphasize that 
converged nonlinear simulations require well-resolved
initial conditions.  One limitation of these studies, however, is that \citet{Robertson2010} and \citet{McNally2012} only provide converged reference simulations for the linear (and possibly weakly nonlinear) phase. 
In addition,  converged nonlinear solutions require solving dissipative equations. Many available astrophysical codes do not implement this essential feature. As a result, these works could only follow the instability for a few e-folding timescales.

Not all works take the benchmark approach, however.  In place of a nonlinear reference solution, some authors use apparent small-scale structure as a proxy for the accuracy of their simulations \citep[e.g.,][]{Springel2010,Hopkins2015}. Presumedly, more small-scale structure implies less numerical dissipation, and therefore greater accuracy. We find in the current paper that this intuition can, in some cases, lead to false conclusions. Some tests also abandon the smooth initial conditions of \citet{Robertson2010} and \citet{McNally2012}, even though this choice precludes convergence of even the linear phase of the instability because the linear growth rates increase with wavenumber for an initially discontinuous velocity profile.

In this paper, we extend the work of \citet{McNally2012} by providing
reference solutions for the \textit{strongly nonlinear} evolution of the
KH instability.  We use a smooth initial condition and
explicit diffusion. We conduct simulations using both Athena (a
Godunov code), and Dedalus (a pseudo-spectral code that can solve the Navier-Stokes equations of compressible hydrodynamics) and find that both converge to the
same reference solutions.  We see agreement among different codes and
different resolutions, with the validity of the reference solution
limited only by (unavoidable) chaotic evolution at late times.  We
propose that future code tests include this KH instability problem and compare to our validated, converged, reference solutions.

We organize the remainder of the paper as follows.  Section~\ref{sec:method}
describes the equations, initial conditions, and codes used for our
simulations.  The results comprise two sections.  In section~\ref{subsec:unstratified-results}
we discuss the simpler simulations with constant initial density.
Section~\ref{sec:drat 2} discusses the more complicated simulations with an initial
density jump. Section~\ref{sec:conclusion}, summarizes our results.

\section{Methods}
\label{sec:method}
\subsection{Equations and Initial Conditions}
\label{subsec:setup}
We solve the hydrodynamic equations, including explicit terms for the diffusion of momentum and temperature:
\begin{subequations}\label{eqn:equations of motion}
\begin{align}
  \frac{\partial \rho}{\partial t}
  + \grad\dot\left(\rho\,  \vec{u}\right) &= 0, \\
  \frac{\partial}{\partial t}(\rho\,\vec{u})
  + \grad\dot\left(P\,\boldsymbol{\mathsf{I}} 
    + \rho\,\vec{u}\otimes\vec{u} \right) &= 
  -\grad\dot\boldsymbol{\mathsf{\Pi}},\\
  \frac{\partial E}{\partial t} 
  + \grad\dot\left[(E+P)\,\vec{u} \right]
  &= \grad\dot(\chi\rho\grad T)
  - \grad\dot(\vec{u}\dot\boldsymbol{\mathsf{\Pi}}),
\end{align}
\end{subequations}
along with the nondimensionalized ideal gas equation of state, $P=\rho T$, with constant ratio of specific heats $\gamma=5/3$.  $\boldsymbol{\mathsf{I}}$ is the identity tensor, $\chi$ is the thermal diffusivity (with units $\text{cm}^2/\text{s}$; $K=n k_{\text{b}} \chi$ is the thermal conductivity), and
\begin{align}
  \boldsymbol{\mathsf{\Pi}} = - \nu \rho \left( \grad \vec{u} + (\grad \vec{u})^T - \frac{2}{3} \boldsymbol{\mathsf{I}} \grad \dot \vec{u}\right)
\end{align}
is the viscous stress tensor with viscosity $\nu$ (with units $\text{cm}^2/\text{s}$).  We assume both $\nu$ and $\chi$ are constant.

We add a passive scalar to our simulations which we refer to as ``dye.''
The local fraction of dye particles $c$ expresses dye concentration, and initially ranges from 0 to 1. 
The local conservation of dye is then
\begin{align}
  \frac{\partial}{\partial t}\left(\rho c\right)
  + \nabla\cdot\left(\rho c\,\vec{u}\right)
  = \rho \frac{d c}{d t} 
  &= -\nabla\cdot\vec{Q}_{\text{dye}},\label{eq:dye-evolution}\\ 
  \vec{Q}_{\text{dye}} &= - \rho \nu_{\text{dye}} \nabla c
\end{align}
where $d/dt$ represents the Lagrangian derivative, and $\nu_{\text{dye}}$ represents a diffusion coefficient for dye molecules (with units $\text{cm}^2/\text{s}$).  These equations conserve the total dye mass $\int{}\rho\,c\,\mathrm{d}V$.

We define a dye entropy per unit mass $s\equiv-\,c\,\ln{}c$, along with its volume integral
\begin{align}\label{eqn:entropy}
S\equiv\int{}\rho\,s\,\mathrm{d}V.
\end{align}
These evolve such that:
\begin{align}
  \rho \frac{d s}{d t} - \nabla\cdot\left[(1+\ln{}c)\,\vec{Q}_{\text{dye}}\right]
  &= \rho\nu_{\text{dye}} \frac{|\nabla c|^2}{c} \label{eq:s-evolution} \\
  \frac{d S}{d t} 
  &= \int \rho \nu_{\text{dye}} \frac{|\nabla c|^2}{c} \mathrm{d}V
  \geq 0.
\end{align}
The second term on the left-hand side of equation~\ref{eq:s-evolution} represents the entropy flux due to reversible diffusion of the dye. The right-hand side represents entropy generation due to non-reversible dissipation.\footnote{Equation~\ref{eq:s-evolution} can be made to look like the analogous equation for heat conduction with the definition of a new ``temperature'' $T_{\text{dye}} \equiv -\frac{1}{1+\ln{}c}$}
The volume-integrated entropy $S$ satisfies the following important properties:
\begin{enumerate}
\item A fully unmixed fluid with $c=0$ or $c=1$ everywhere has zero entropy ($S=0$).
\item A fully mixed fluid with $c^*=\int{}\rho\,c\,\mathrm{d}V/\int{}\rho\,\mathrm{d}V$ maximizes the entropy, $S_{\text{max}}=-c^{*}\ln{}c^*\,\int{}\rho\,\mathrm{d}V$.
\item $S$ increases monotonically with time if $\nu_{\text{dye}} > 0$, and stays constant otherwise.
\end{enumerate}

We restrict our attention to periodic simulations.
This avoids potential difficulties with imposing Dirichlet and/or Neumann boundary conditions.  Our initial conditions are:
\begin{subequations}\label{eqn:ICs}
\begin{align}
  \rho &= 1 +
  \frac{\Delta\rho}{\rho_0}
  \times\frac{1}{2}\left[\tanh\left(\frac{z-z_1}{a}\right) - \tanh\left(\frac{z-z_2}{a}\right)\right]\label{eq:initial-rho}\\
  u_x &= u_{\text{flow}} \times \left[\tanh\left(\frac{z-z_1}{a}\right) - \tanh\left(\frac{z-z_2}{a}\right) - 1 \right] \\
  u_z &= A \sin(2\pi{}x) \times \left[\exp\left(-\frac{(z-z_1)^2}{\sigma^2}\right) + \exp\left(-\frac{(z-z_2)^2}{\sigma^2}\right)\right] \\
  P &= P_0 \\
  c &= \frac{1}{2}\left[\tanh\left(\frac{z-z_2}{a}\right) - \tanh\left(\frac{z-z_1}{a}\right) + 2\right],
\end{align}
\end{subequations}
where $a=0.05$ and $\sigma=0.2$ are chosen so that the initial condition is resolved in all of our simulations.  We take $u_{\text{flow}}=1$ and $P_0=10$ so that the flow is subsonic with a Mach number $M\sim{}0.25$.  The size of the initial vertical velocity perturbation is $A=0.01$.  The Athena simulations are initialized with these functions evaluated at cell-centers even though Athena data represents cell-averaged quantities (see Appendix~\ref{sec:interpolate} for more discussion of this effect).

We adopt a rectangular domain with $x$ in $[0,L_x)$, and $z$ in $[0,2L_z)$, with $L_x = 1$ and $2 L_z = 2$, and $z_1=0.5$, $z_2=1.5$, with periodic boundary conditions in both directions.  The simulations have a horizontal resolution of $N$ grid points (in Athena) or modes (in Dedalus) in the $x$ direction, and $2N$ grid points/modes in the $z$ direction.  Our initial condition has a reflect-and-shift symmetry: taking $z\rightarrow 2-z$ and $x\rightarrow x + 1/2$ changes the sign of $u_z$ but leaves the other quantities invariant.  Thus, the simulations solve for the same flow twice.  This is a requirement when using periodic boundary conditions, but also provides a test of whether or not the numerical simulations can preserve the symmetry. Almost all simulations presented here maintain the symmetry. We therefore only show the lower half of the domain.  We calculate volume-averaged quantities like the dye entropy or the $L_2$ norm with respect to the entire domain.

In equation~\ref{eq:initial-rho}, the free parameter $\Delta \rho/\rho_0$ represents the density jump across the interface.
We study simulations with $\Delta \rho/\rho_0 = 0$ in section~\ref{subsec:unstratified-results} and with $\Delta \rho/ \rho_0=1$ in section~\ref{sec:drat 2}.  We refer to this change in density as a ``jump'' throughout, although the transition is smooth, set by the tanh in equation~\ref{eq:initial-rho}.
The Reynolds number $\text{Re}$ quantifies diffusion,
\begin{align}\label{eqn:Re}
  \nu = \chi = \nu_{\text{dye}} = \frac{L \Delta u}{\text{Re}},
\end{align}
where $\Delta u = 2u_{\text{flow}}$ is the change in velocity.
Note that we set the thermal diffusivity $\chi$ constant; consequently, the thermal conductivity $K \propto \rho$.  Throughout the paper we measure time in units of $L/u_{\text{flow}}$, so $t=1$ corresponds to approximately one turnover time.
Equations~\ref{eqn:equations of motion}--\ref{eqn:Re} specify our system, with the free parameters $\Delta \rho/\rho_0$ and ${\rm Re}$.  In the following section we detail our methods for solving this system of equations.

\subsection{Numerical Methods}
\label{sec:setup}
We study the KH instability using two open-source codes employing very different numerical methods:  Athena \& Dedalus.

Athena\footnote{Athena is available at \url{https://trac.princeton.edu/Athena/}.} is a finite-volume Godunov code \citep{gs08,stone08}.  The scheme represents all field quantities with volume averaged values in each grid element. A Riemann problem solves for fluxes between elements.  We use third-order reconstruction with limiting in the characteristic variables to approximate field values at the element walls, the HLLC Riemann solver, and the CTU integrator.  We used the ``-O3'' compiler flag using Intel 14.0.1.106 and Mvapich2 2.0b on the Stampede supercomputer.  We repeated some runs using second-order reconstruction and/or the Roe Riemann solver and/or stricter compiler flags (e.g., ``-O2 -fp-model strict'') --- these choices did not qualitatively affect the solutions.  We use a static, uniform mesh, and a CFL safety factor of $0.8$.

Athena is second-order accurate in both space and time.  The leading-order grid-scale errors are diffusive.  For most simulations reported here, we include explicit diffusion.  A sufficiently large explicit diffusion can dominate grid-scale errors and allow the simulation to remain close to the true solution.  However, higher-order grid-scale errors can introduce non-diffusive effects, such as dispersion.  If higher-order errors project onto unstable modes, they can cause large differences in the solution, despite being higher order.  The grid-scale errors in Athena respect the reflect-and-shift symmetry of our problem up to floating point accuracy, so even non-converged simulations can maintain the initial symmetry of the flow.  In practice, we find all simulations maintain the initial symmetry, except simulations with $\Delta\rho/\rho_0=1$ without explicit diffusion.  Since Athena's algorithm manifestly preserves this symmetry, we expect the error results from chaotic amplification of floating-point errors.

Dedalus\footnote{Dedalus is available at \url{http://dedalus-project.org}.} is a pseudo-spectral code \citep{burns16}.  All field variables are represented as Fourier series, and the simulation solves for the evolution of the spectral-expansion coefficients in time.  The code evaluates nonlinear terms on a grid with a factor 3/2 more points than Fourier coefficients; i.e., the 2/3 de-aliasing rule.   \citet{lecoanet14} (appendix D.1) describes our implementation of the Navier-Stokes equations.  Our implementation of the dye evolution equation is
\begin{subequations}
\begin{align}
  \partial_t c - & \nu_{\text{dye}} \left(\partial_x^2 c + \partial_z c_z\right) = \nonumber \\
   & \quad \quad - u\partial_x c  - w c_z + \nu_{\text{dye}}\left(\partial_x\Upsilon'\partial_x c + \partial_z\Upsilon' c_z\right), \\
  & c_z - \partial_z c = 0,
\end{align}
\end{subequations}
where we use the same notation as \citet{lecoanet14}.  For timestepping, we use a third-order, four-stage DIRK/ERK method (RK443 of \citealt{ascher97}) with a total CFL safety factor of 0.6 (i.e., 0.15 per stage).  This formulation allows implicit timestepping of sound waves.  Thus, our timestep size only adjusts with the flow velocity, not the sound speed.  The excellent agreement between the highest resolution Dedalus and Athena simulations shows that high-wavenumber sound waves have negligible influence on the solution.  

The pseudo-spectral method produces almost no numerical diffusion. Stability concerns require explicit diffusion in nonlinear calculations.  In marginally resolved simulations, discretization errors manifest as Gibbs' ringing, which is prominently visible in snapshots.  The numerical method does not explicitly preserve the reflect-and-shift symmetry---numerical errors can put power into the asymmetric modes.  However, we find that in resolved simulations these asymmetric modes never grow to large amplitudes.  Thus, maintaining this symmetry gives a test for a simulation's fidelity.

\section{Results}
\label{sec:results}

This section describes the nonlinear evolution of the KH instability, provides reference solutions, and compares the performance of Dedalus and Athena.  Section~\ref{subsec:unstratified-results} considers unstratified simulations with constant initial density; both codes handle this problem easily.  Section~\ref{sec:drat 2} concerns simulations with a density jump across the shear interface.  This problem shows rich behavior and poses significant numerical challenges.

\subsection{Unstratified simulations ($\Delta\rho/\rho_0 = 0$)}
\label{subsec:unstratified-results}

\begin{figure*}
\includegraphics[width=\textwidth]{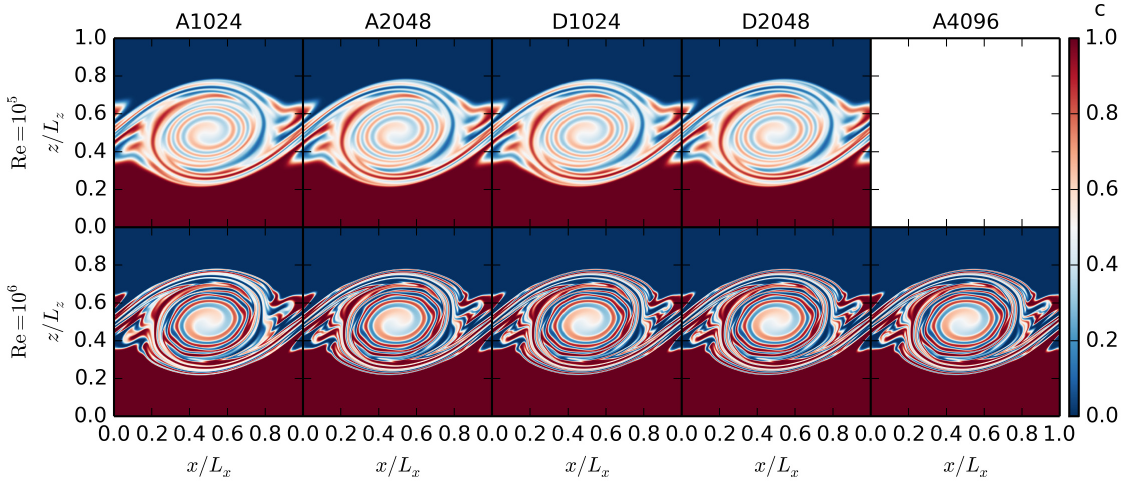}
  \caption{Snapshots of the dye concentration field in several simulations with $\Delta\rho/\rho_0=0$ at $t=6$.  The upper (lower) row shows simulations with ${\rm Re}=10^5$ ($10^6$).  All the simulations with ${\rm Re}=10^5$ are well resolved.  Small differences exist between the lower-resolution Athena simulations at ${\rm Re}=10^6$ and the highest-resolution Athena simulation \& Dedalus simulations (e.g., near $(x,z)=(0.9,0.6)$, see Figure~\ref{fig:zoomin}).}\label{fig:unstratified-dye}
\end{figure*}

In this section, we discuss simulations with constant initial density ($\Delta \rho/\rho_0=0$).  Figure~\ref{fig:unstratified-dye} visualizes the flow with the dye concentration field of the lower half of the domain for simulations with explicit diffusion at different resolutions and Reynolds number, $\text{Re}$.  The snapshots show the state at $t=6$.  Strong nonlinearity begins at $t\sim 2$, so this corresponds to at least four turnover times after the initial saturation of the instability.  The simulations are labeled by the code used (A for Athena; D for Dedalus), and their horizontal resolution.

The flow consists of coherent filaments of unmixed fluid with dye concentration close to zero or one.  The filaments twist around the central vortex until they become thin enough to diffuse away.  The central vortex stays coherent in all simulations, and exhibits a more gradual dye-concentration gradient than in the filaments.  This reflects the smooth velocity and dye initial condition.

\subsubsection{${\rm Re}=10^5$}

Many of the simulations with the same $\text{Re}$ but different resolution look similar by eye.  To more quantitatively assess convergence, we calculate the $L_2$ norm of the differences between dye concentration fields in different simulations:
\begin{align}
L_2(c_{\rm X} - c_{\rm Y}) = \left[\int \mathrm{d}V \ (c_{\rm X}-c_{\rm Y})^2\right]^{1/2},
\end{align}
where $c_{\rm X}$ and $c_{\rm Y}$ represent the dye concentration fields in two simulations, X and Y.  The Athena and Dedalus grids are different, so we use spectrally accurate techniques to interpolate Dedalus solutions to the Athena grid for direct comparison (Appendix~\ref{sec:interpolate}).  We argue in Appendix~\ref{sec:convergence} that all simulations converge to our highest-resolution Dedalus simulations; thus, we assume these simulations are a good approximation to the ``true'' solution.

\begin{figure}
\includegraphics[width=\columnwidth]{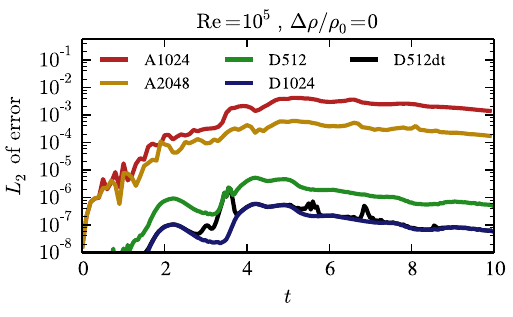}
  \caption{$L_2$ norm of dye-concentration errors for $\Delta\rho/\rho_0=0$ and ${\rm Re}=10^5$.  We take  D2048 as the ``true'' solution (see Appendix~\ref{sec:convergence}). Both Dedalus and Athena exhibit third-order convergence.  D512dt is run with half the timestep size as D512.  Its error is similar to D1024, showing that the higher accuracy of D1024 is mostly due to a smaller timestep size rather than higher spatial resolution.}\label{fig:error_1e5_1}
\end{figure}

Figure~\ref{fig:error_1e5_1} shows the $L_2$ norm of the difference between dye concentration fields of D2048 and other simulations with ${\rm Re}=10^5$.  Because we believe D2048 closely represents the true solution (Appendix~\ref{sec:convergence}), we call this the $L_2$ norm of the error.  Solutions from both codes approach D2048 as resolution increases.  At late times, A2048 and D1024 have roughly eight-times smaller errors than A1024 and D512, respectively.  That is, both codes exhibit third-order convergence.  This indicates that interpolation produces the dominant error in  Athena, which is the only third-order part of the algorithm.  The Dedalus simulations are spatially resolved, so timestepping produces the  dominant error source in the Dedalus simulations, which is also third order.  We also plot errors from D512dt, which is run with a horizontal resolution of 512, but with half the CFL safety factor.  D512dt is almost as accurate as D1024, showing that the higher accuracy of D1024 is mostly due to taking smaller timesteps.  There are certain times (most notably near $t=3.5$) where the flow develops smaller structures, and extra spatial resolution is required.  The errors in quantities other than dye concentration (e.g., density) follow similar behavior to that shown in Figure~\ref{fig:error_1e5_1}.

\begin{figure}
 \includegraphics[width=\columnwidth]{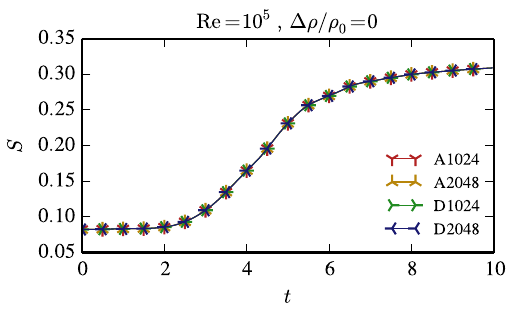}
  \caption{Volume-integrated dye entropy (equation~\ref{eqn:entropy}) as a function of time for the four simulations with ${\rm Re}=10^5$ shown in Figure~\ref{fig:unstratified-dye}.  All simulations are well resolved, so the dye entropies are almost equal.}\label{fig:entropy_1e5_1}
\end{figure}

We calculate the volume-integrated dye entropy for each simulation (equation~\ref{eqn:entropy}).  Figure~\ref{fig:entropy_1e5_1} plots the entropy as a function of time.  Because all simulations are well resolved, there are no visible differences in the entropy between the different simulations.

\subsubsection{${\rm Re}=10^6$}\label{sec:1e6}

The unmixed filaments are much thinner for ${\rm Re}=10^6$ than for ${\rm Re}=10^5$, challenging  the codes.  Unlike the ${\rm Re}=10^5$ case, some minor visible differences appear between the solutions for ${\rm Re}=10^6$. The lower-resolution simulations do not fully resolve the flow (one such feature is highlighted in Figure~\ref{fig:zoomin}).

\begin{figure}
 \includegraphics[width=\columnwidth]{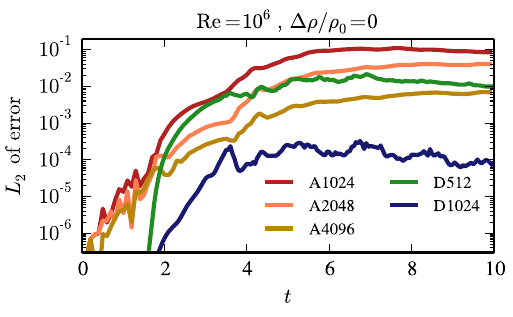}
  \caption{$L_2$ norm of dye-concentration errors for $\Delta\rho/\rho_0=0$ and ${\rm Re}=10^6$.  A1024 is not well resolved so its errors follow a different pattern than the other Athena simulations.  The errors in A4096 are smaller than the errors in A2048 by $\approx 6$.  The errors in D1024 are smaller than the errors in D512 by about 100.  This demonstrates the fast (exponential) convergence of spectral methods.}\label{fig:error_1e6_1}
\end{figure}

To assess convergence, we again plot the $L_2$ norm of the error in dye concentration with respect to D2048 (Figure~\ref{fig:error_1e6_1}).  A1024 has the largest errors of any simulation. At late times, the errors interact nonlinearly, whereas the errors in the higher-resolution Athena simulations stay linear and the temporal variation of the error is the same independent of the magnitude of the error.  The ratio of errors of the two higher-resolution Athena simulations is about $6$---in between second- and third-order convergence.  This suggests that the size of interpolation errors roughly match the size of other errors in the code (e.g., from the Riemann problem or timestepping).

The difference in errors between D512 and D1024 is about 100---much larger than the difference in errors between the Athena simulations. D512 (not shown in Figure~\ref{fig:unstratified-dye}) under resolves the flow and includes some low-amplitude Gibbs' ringing. Increasing the resolution from 512 to 1024 eliminates spatial errors because of the exponential convergence of spectral methods.  This allows for very large error reduction with only modest resolution changes.  The exponential nature of spectral methods makes convergence practically binary:  simulations with Gibbs' ringing are not converged; simulations without Gibbs' ringing very likely are converged.

\begin{figure}
 \includegraphics[width=\columnwidth]{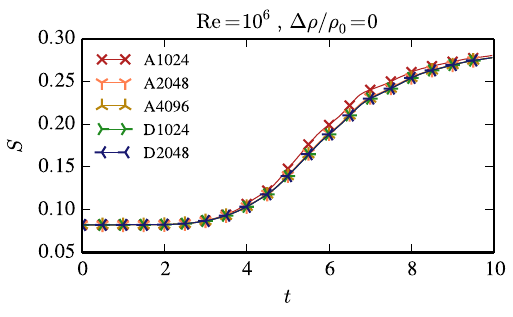}
  \caption{Volume-integrated dye entropy (equation~\ref{eqn:entropy}) as a function of time for the five simulations with ${\rm Re}=10^6$ shown in Figure~\ref{fig:unstratified-dye}.  The entropy of all simulations are very similar except for A1024; this is another indication that A1024 is not well resolved.}\label{fig:entropy_1e6_1}
\end{figure}

We plot volume-integrated dye entropy for ${\rm Re}=10^6$ in Figure~\ref{fig:entropy_1e6_1}.  Like for ${\rm Re}=10^5$, all well-resolved simulations produce similar entropy.  However, the under-resolved A1024 produces slightly more entropy.  This agrees with the heuristic that extra numerical diffusion leads to excess entropy generation.

\subsubsection{An effective Reynolds number?}

We now describe Athena simulations without any explicit diffusion.  An important question is, does the numerical diffusion in Athena act like an explicit diffusion?  Put another way, does Athena have an effective Reynolds number at a given resolution for this problem?  As we describe below and in section~\ref{sec:drat 2}, the answer to this question is very problem dependent.

\begin{figure}
 \includegraphics[width=\columnwidth]{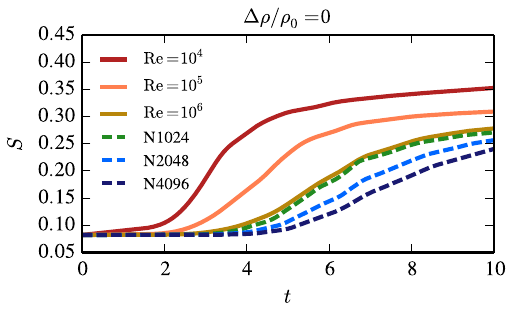}
  \caption{Volume-integrated dye entropy (see section~\ref{sec:setup}) as a function of time with $\Delta\rho/\rho_0=0$, for three resolved simulations with different ${\rm Re}$, as well as three Athena simulations with no explicit diffusion (dashed lines; labeled with N, for no explicit diffusion, and their horizontal resolution).  The entropy of N1024 and the simulation with ${\rm Re}=10^6$ are very similar. Their flow fields show minor differences (see Figure~\ref{fig:zoomin}).  Note that the entropy decreases with increasing resolution in the simulations without explicit diffusion.   This is not the case in simulations with an initial density jump (see Figure~\ref{fig:entropy_1e5_2_nodiff}).}\label{fig:entropy_conv}
\end{figure}

To test this, we plot the converged volume-integrated dye entropy for several Reynolds numbers, along with the volume-integrated dye entropy for Athena simulations without explicit diffusion (Figure~\ref{fig:entropy_conv}).  The entropy evolution of N1024 is similar to the entropy evolution for ${\rm Re}=10^6$.  This might lead one to think that the effective Reynolds number of this Athena simulation is about $10^6$.

\begin{figure}
\includegraphics[width=\columnwidth]{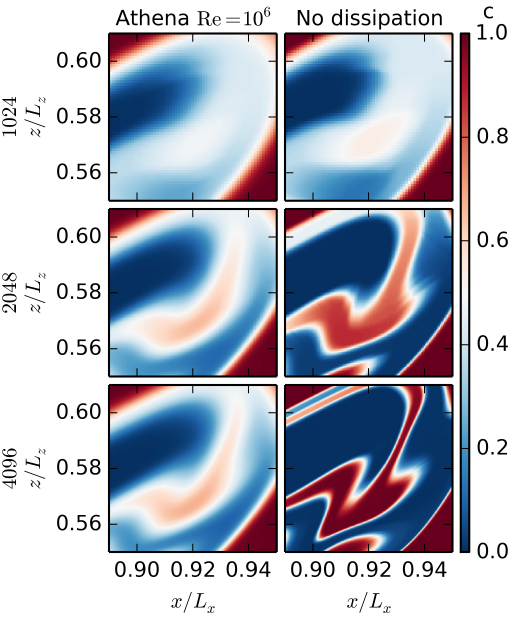}
  \caption{Snapshots of the dye concentration field between $0.89<x<0.95$ and $0.55<z<0.61$, at $t=6$ for $\Delta\rho/\rho_0=0$.  All simulations use Athena, either with ${\rm Re}=10^6$ (left column) or no explicit diffusion (right column).  The three rows have different resolutions.  This zoom-in of Figure~\ref{fig:unstratified-dye} highlights the differences between simulations at different resolutions---however, for the most part, the simulations look very similar.  A2048 \& A4096 represent resolved simulations with ${\rm Re}=10^6$.  Although the entropies for N1024 (upper right plot) \& A4096 (lower left plot) track each other (Figure~\ref{fig:entropy_conv}), the dye concentration fields exhibit minor differences.}\label{fig:zoomin}
\end{figure}

However, a closer investigation shows that N1024 and the ${\rm Re}=10^6$ simulation have different dye concentration fields which, by chance, result in similar volume-integrated entropies (Figure~\ref{fig:zoomin}).  Instead, the dye concentration field of N1024 looks like the dye concentration field of the (under resolved) A1024 simulation with ${\rm Re}=10^6$.  Figure~\ref{fig:entropy_1e6_1} shows A1024 has a higher entropy than the true ${\rm Re}=10^6$ solution.  By removing the explicit diffusion, the flow evolution remains similar to A1024 (and different from the resolved ${\rm Re}=10^6$ solution), but the interfaces between filaments are sharper, which decreases the entropy.  The effects of having the incorrect flow field (increasing entropy), but sharper interfaces between filaments (decreasing entropy) happen to cancel out, so the entropy of N1024 is similar to that of ${\rm Re}=10^6$.

Although we have highlighted the differences between N1024 and the converged solutions with ${\rm Re} = 10^6$, it is worth reiterating that the two solutions are in fact remarkably similar.  This shows that N1024 roughly has an effective Reynolds number of $10^6$. In detail, however, the remaining modest differences between N1024 and the ${\rm Re}=10^6$ solution demonstrate that the numerical dissipation in Athena is not exactly equivalent to physical dissipation via viscosity and thermal conduction.

One difficulty with the notion of an effective Reynolds number is that it is extremely problem dependent, even at fixed resolution.  In the next section,  we introduce a small (by astrophysical standards) density jump into the initial condition.  This completely changes the problem by introducing secondary instabilities which enhance mixing, producing very clear differences between resolved simulations and Athena simulations without explicit diffusion (Figure~\ref{fig:nodiff}).  For the constant-density problem described here, omitting diffusion produces less entropy.  Including a density jump reverses this trend: simulations with only numerical diffusion undergo {\it more} mixing than simulations with explicit diffusion.  Although assigning an effective Reynolds number to Athena simulations without explicit diffusion may be reasonably accurate for the constant-initial-density problem, this does not carry over to the problem with an initial density jump.

\begin{figure*}
\includegraphics[width=\textwidth]{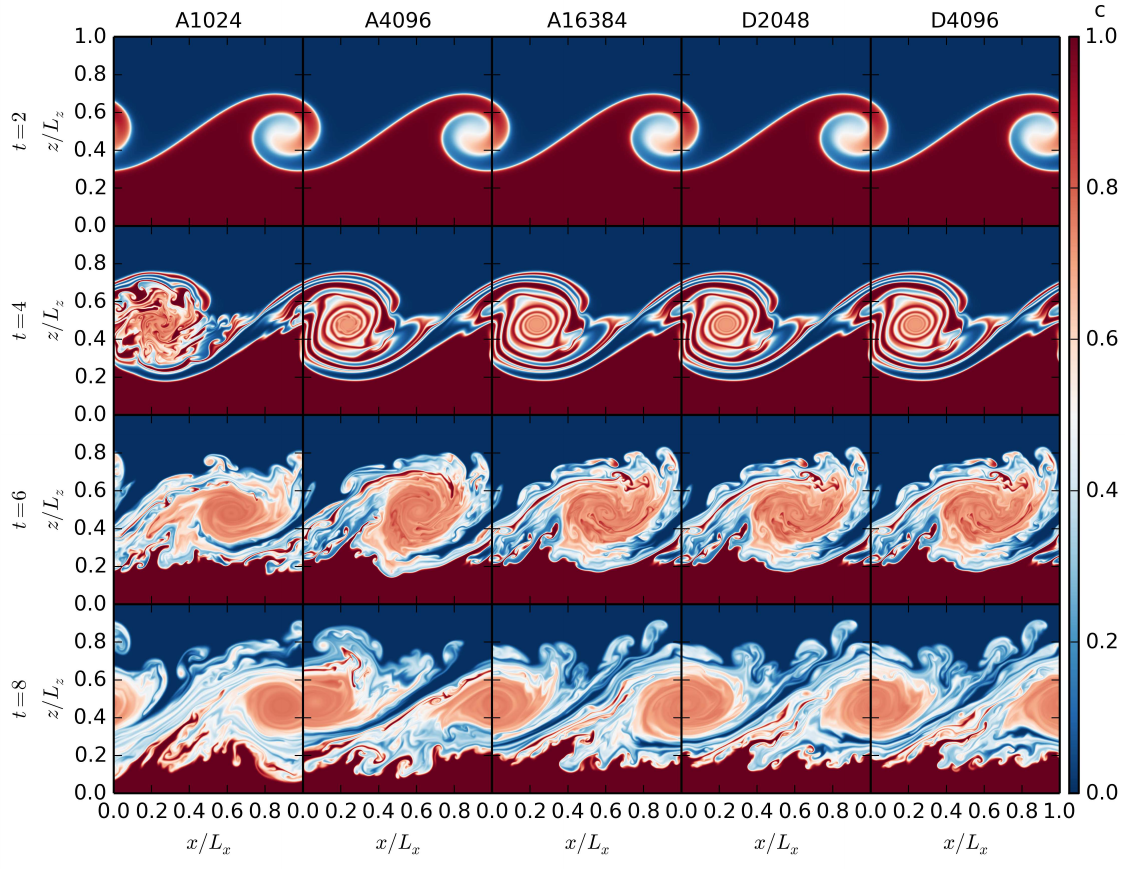}
  \caption{Snapshots of the dye concentration field in several simulations with $\Delta\rho/\rho_0=1$ and ${\rm Re}=10^5$.  Each row corresponds to a different time.  The low-resolution Athena simulations suffer from a secondary instability (seen at $t=4$) in the middle of the vortex, which is not present in the Dedalus simulations nor A16384.  This causes substantial differences at later times.  A16384 and both Dedalus simulations stay very similar at late times, although small differences develop from chaos (see section~\ref{sec:chaos}).}\label{fig:stratified-dye}
\end{figure*}

\subsection{Simulations with a density jump ($\Delta\rho/\rho_0 = 1$)}\label{sec:drat 2}

Both the qualitative features of the flow and the convergence properties of the simulations change dramatically once we introduce an initial density jump ($\Delta \rho/\rho_0\neq 0$).  Unlike the unstratified case, secondary instabilities of the filaments produce small-scale structures in the flow.  These secondary instabilities, and the resulting small-scale features, depend on the resolution and the code used.  As a result, simulations with a nonzero density jump require far more computational resources than the unstratified simulations presented in the previous section.  We limit the simulations with explicit diffusion to ${\rm Re}=10^5$---our finite-computing budget precludes solutions for ${\rm Re}=10^6$. The largest simulations required roughly $10^6$ core-hours. 

\begin{figure*}
\includegraphics[width=\textwidth]{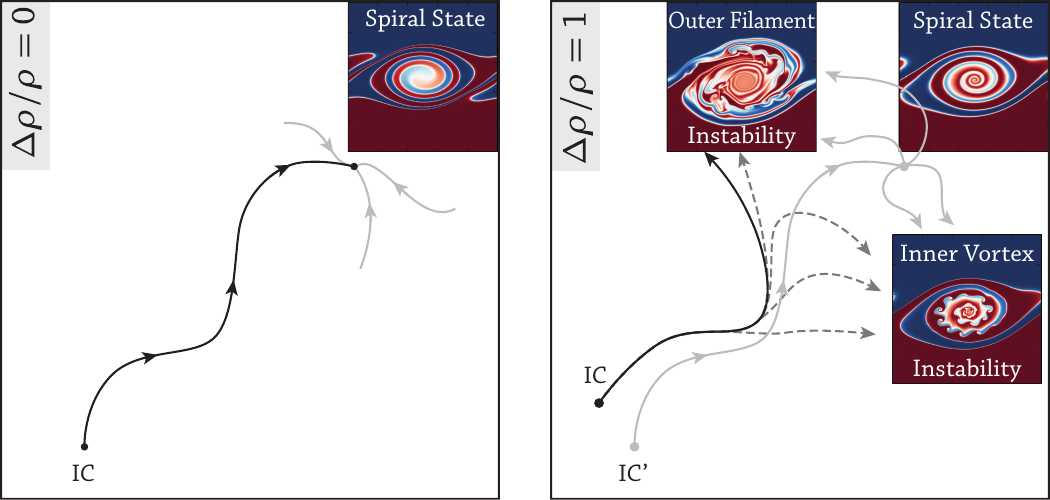}
  \caption{Schematic phase-space diagram for $\Delta\rho/\rho_0=0$ (left) and $\Delta\rho/\rho_0=1$ (right).  For constant initial density, the system has a stable state with ever-narrowing spiral filaments.  We hypothesize that there is an initial condition IC' (right panel)  leading to a similar spiral state for $\Delta\rho/\rho_0=1$.  But this state is now unstable to the outer filament instability (OFI) and the inner vortex instability (IVI).  Our chosen initial condition's (IC) trajectory (solid black line) approaches the spiral state, but becomes unstable to the OFI.  Errors introduced by the numerical hydrodynamics may cause deviations in the trajectory leading to the IVI (dashed grey lines).}\label{fig:schematic}
\end{figure*}

Figure~\ref{fig:stratified-dye} shows the dye concentration for different simulations at different times.  In both Dedalus simulations, and the highest-resolution Athena simulation, the outer filaments (i.e., those outside the central vortex) become unstable to a sausage-like mode (see the panel in Figure~\ref{fig:schematic} for an example).  Lower-resolution Athena simulations also undergo a separate instability of the inner filaments of the vortex.  We refer to these two instabilities at the outer-filament instability (OFI) and the inner-vortex instability (IVI) (see Figure~\ref{fig:schematic} for examples).  These instabilities are similar to the baroclinic secondary instabilities discussed in \citet{Reinaud2000,Fontane2008}.  The competition between these two instabilities plays a crucial role in the evolution of the system.

\begin{figure}
 \includegraphics[width=\columnwidth]{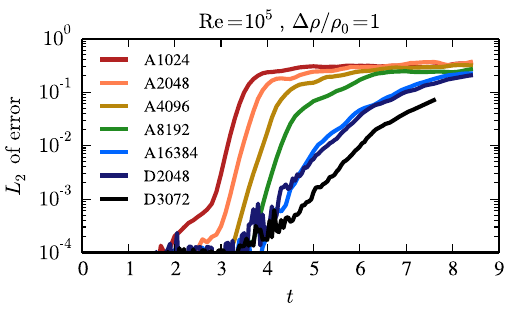}
  \caption{$L_2$ norm of dye-concentration errors for $\Delta\rho/\rho_0=1$ and ${\rm Re}=10^5$.  D3072 and D4096 are the closest pair of simulations, suggesting that D4096 is a good approximation to the true solution.  All Athena simulations except A16384 diverge away from D4096 exponentially with a rate of 8, suggesting the growth rate of the inner vortex instability (see Figure~\ref{fig:schematic}) is also 8.  The errors in the lower-resolution Dedalus simulation and A16384  grow exponentially with a rate of about 2-3.  We interpret this divergence as due to chaos (see section~\ref{sec:chaos}).  D3072 has errors smaller than D2048 by $\approx 4$, consistent with third-order convergence set by our choice of timestepping algorithm.}\label{fig:diff_1e5_2}
\end{figure}

We plot the $L_2$ norm of the error in dye concentration with respect to D4096 in Figure~\ref{fig:diff_1e5_2}.  As described in Appendix~\ref{sec:convergence}, we believe D4096 approximates the true solution. The difference between D3072 and D4096 are smaller than the differences between any other pair of simulations.  At later times, even the errors between D3072 and D4096 become large.  In section~\ref{sec:chaos} we attribute this late-time behavior to chaos.

Figure~\ref{fig:diff_1e5_2} shows that at early times, the low-resolution Athena simulations diverge exponentially from D4096 with an inferred  growth rate of about 8.  The IVI produces this divergence.  Furthermore, the four Athena simulations with resolutions between 1024 and 8192 are all equally spaced horizontally in Figure~\ref{fig:diff_1e5_2}.  The horizontal-axis spacing is $\log{2}/2$ time units.  This suggests that the same instability exists independent of resolution, but the amplitude of the perturbation that seeds the instability drops by 16 when the resolution doubles.  Though numerical errors seed the growth, the constant growth rate of the IVI suggests it is a physical instability (we demonstrate this in section~\ref{sec:IVI}).

The IVI is a robust feature of low-resolution Athena simulations.  Using the Roe integrator, second-order reconstruction, or shifting the initial condition by half a grid point does not affect the development of this instability (as confirmed using the $L_2$ error), but can cause visible differences in the flow evolution.  This demonstrates that grid-scale errors drive the IVI.  Using first-order reconstruction suppresses the IVI, but the enhanced numerical diffusion causes large errors.  We have also tried adding low-amplitude (up to $10^{-4}$) white noise to the initial density or pressure.  These do not cause any visible changes to the IVI.  The flow forgets some of the detailed information of its initial condition (see section~\ref{sec:perturb}).

The highest-resolution Athena simulation (A16384) does not develop the IVI.  This demonstrates that the initial condition is in fact stable to the IVI; the problem is well-posed.  Rather, numerical errors seed the IVI at some later time, during the evolution of the flow.  Although some numerical errors are still inevitably present, A16384 does not develop the IVI because the ``base state'' of spiralling filaments of unmixed fluid also succombs to the OFI.  In this case, the OFI disrupts the inner vortex before the IVI grows to large amplitudes (see Figure~\ref{fig:schematic}).  

The absence of the IVI is a robust feature of our Dedalus simulations.  We confirmed the stability of the base state by re-running D2048 with low-amplitude white noise added to the initial condition; we also re-initialized D2048 from the Athena initial condition. This introduces small but non-random grid-representation differences (section~\ref{sec:IVI}).  In both cases, we recover the same evolution.  However, we can trigger the IVI in Dedalus with a large ($\sim 10\%$ by energy) perturbation to the initial condition (section~\ref{sec:perturb}).

Figure~\ref{fig:schematic} summarizes the relation between the two secondary instabilities in this problem.  For a constant initial density (left panel), the system evolves toward a stable state characterized by spiraling filaments.  Small differences in initial conditions, integration algorithms, presence of dissipation, etc., cause only minor changes in the evolution.  We hypothesize that a similar spiral state also exists for $\Delta\rho/\rho_0=1$, and that it could be reached from some initial condition IC'.  However, our simulations demonstrate that the spiral state is now unstable.  Thus, small errors lead to the large differences in evolution.

Small perturbations to the hypothetical IC' of Figure~\ref{fig:schematic} would lead to trajectories that either develop the OFI or the IVI.  However, our chosen initial condition, IC, is squarely in the attracting basin of the OFI.  Thus, infinitesimal perturbations to IC will still lead to the OFI.  Errors introduced by numerical hydrodynamics cause the codes to not follow the correct trajectory (solid black line).  Certain types of errors can cause trajectories to diverge from the correct solution, sometimes toward the IVI (dashed grey lines).  Alternatively, sufficiently large initial perturbations can also knock the system into the attracting basin of the IVI (section~\ref{sec:perturb}).

We note that the phase space for this problem is very high dimension, and that the outer filament instability and inner vortex instability represent two (likely non-parallel) unstable directions of the spiral state's stable manifold.  Thus, both instabilities can act simultaneously, which sometimes occurs in simulations.  

\begin{figure}
 \includegraphics[width=\columnwidth]{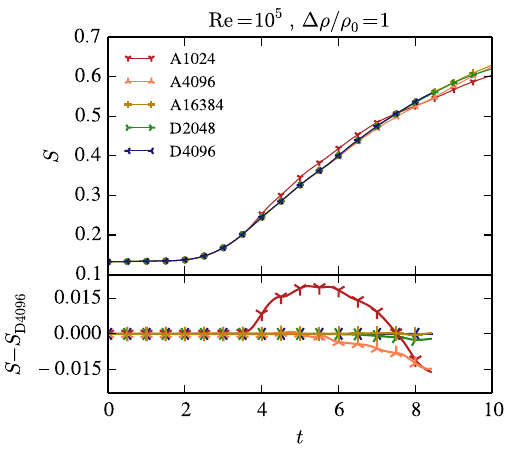}
  \caption{Volume-integrated dye entropy (equation~\ref{eqn:entropy}) as a function of time for simulations with $\Delta\rho/\rho_0 = 1$ and ${\rm Re}=10^5$.  The top panel plots the entropy, and the bottom panel plots the entropy deviation from D4096.  The entropy of all the simulations diverge from D4096, but the less-accurate simulations diverge faster.  For each Athena simulation, the entropy initially increases faster than D4096 when it starts to diverge.  At later times, the entropy sometimes drops below the entropy of D4096.}\label{fig:entropy_1e5_2}
\end{figure}

Figure~\ref{fig:entropy_1e5_2} shows the volume-integrated dye entropy of the simulations shown in Figure~\ref{fig:stratified-dye}.  The entropy follows a similar evolution in every simulation.  To visualize the small deviations, the bottom panel shows the entropy with reference solution D4096 subtracted off.  All the simulations diverge from D4096, but more accurate simulation diverge later, with D2048 and A16384 developing small differences later than any other simulation.  The relation between entropy and resolution is more complicated for $\Delta \rho/\rho_0 = 1$ than for $\Delta \rho/\rho_0 = 0$ (Figures~\ref{fig:entropy_1e5_1} \& \ref{fig:entropy_1e6_1}).

\begin{figure}
\includegraphics[width=\columnwidth]{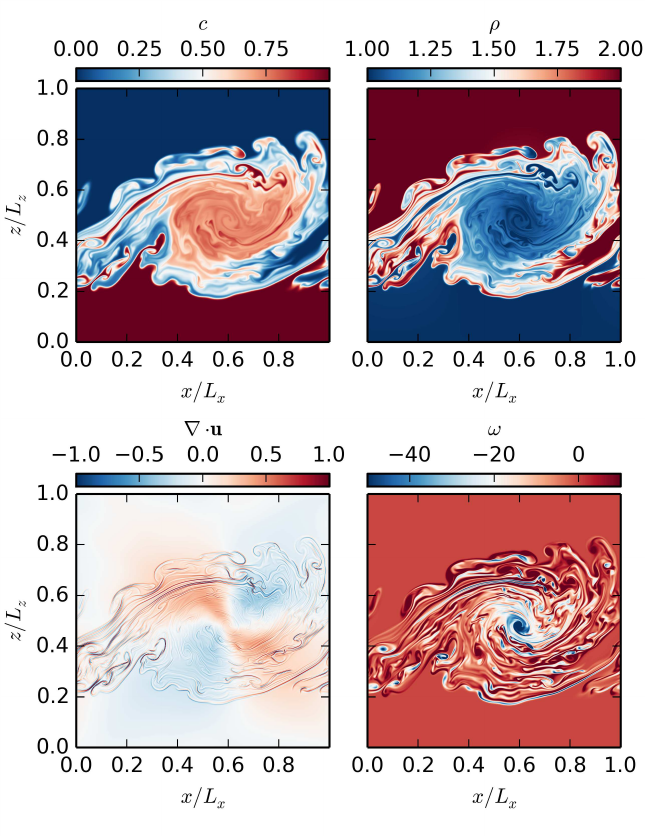}
  \caption{Plots of dye concentration ($c$), mass density ($\rho$), the divergence of the velocity ($\grad\dot\vec{u}$), and the vorticity ($\omega=\vec{e}_z\dot\grad\vec{\times}\vec{u}$) in D4096 with $\Delta\rho/\rho_0 =1$, ${\rm Re}=10^5$ at $t=6$.  The divergence of the velocity and the vorticity are measured in units of $u_{\rm flow}/L_x$.  The dye concentration and mass density fields are almost inverses of each other.  The divergence of the velocity is largest at the interfaces between filaments, whereas the vorticity shows the location of vortices.}\label{fig:fields}
\end{figure}

Apart from the dye concentration field, many of the other flow quantities follow similar patterns. Figure~\ref{fig:fields} shows several quantities from D4096 at $t=6$.  The mass density is almost the inverse of the dye concentration.  This indicates that compression is not an important part of the large-scale dynamics.  Lacking mass diffusion, the density shows sharper gradients than the concentration field.  Temperature diffusion and rapid sound waves regularize the density evolution. These effects limit large temperature gradients, and keep the flow in local pressure equilibrium.  

The velocity divergence field is characterized by a large scale quadrupole centered at the vortex, and large amplitude, small scale features near the boundaries of filaments. The most prominent feature of the vorticity field is the central vortex, which is a remnant of the initial shear.  Small-scale vortex sheets and filaments perhaps result from the incomplete roll-up of the initial condition due to secondary instabilities.

Throughout this paper, we compare different solutions by calculating the $L_2$ norm of the difference between dye concentration fields.  We have made similar comparisons between simulations with ${\rm Re}=10^5$ and $\Delta\rho/\rho_0=1$ using the $L_1$ norm of the difference between dye concentration fields, and using the $L_2$ norm of the difference between the three other fields shown in Figure~\ref{fig:fields}.  We find the results to be qualitatively similar in all cases.  This is expected given the similarity between the fields.

\subsubsection{Inner-vortex instability}\label{sec:IVI}

\begin{figure}
\includegraphics[width=\columnwidth]{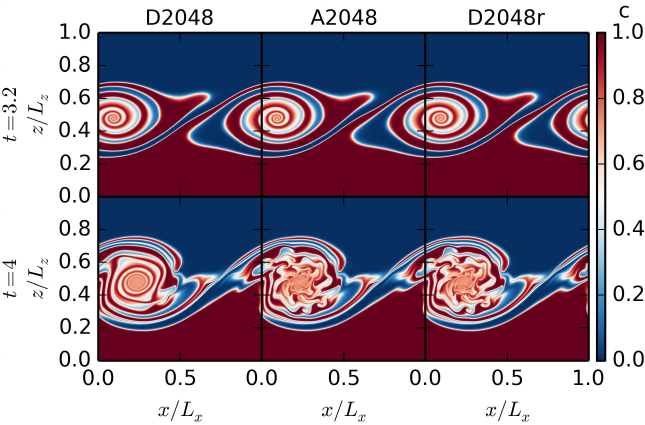}
  \caption{Snapshots of dye concentration field for ${\rm Re}=10^5$ and $\Delta\rho/\rho_0=1$.  D2048r is a Dedalus simulation restarted with the A2048 output at $t=3.2$.  At this time, the inner vortex instability is still in the linear phase, so there are no visible differences between the three simulations.  At $t=4$, the IVI is very nonlinear, producing large differences between D2048 and A2048.  This instability also takes place in D2048r, and the dye concentration fields of A2048 and D2048r are nearly identical.  This demonstrates that the IVI is physical, but is seeded by errors in the lower-resolution Athena simulations that are not present in the Dedalus simulations or the highest-resolution Athena simulations.}\label{fig:IVI}
\end{figure}

To determine the origin (physical vs numerical) of IVI, we initialize a Dedalus simulation with horizontal resolution 2048 with the output from A2048 at $t=3.2$.  We call this simulation D2048r.  Figure~\ref{fig:diff_1e5_2} shows that A2048 is still in the linear phase of the IVI at this time.  In Figure~\ref{fig:IVI}, we plot the dye concentration field at $t=3.2$ and $t=4$ for D2048, A2048, and D2048r.  At $t=3.2$, the simulations all look the same. However, the instability becomes nonlinear by $t=4$,  producing large changes in the dye concentration field.  D2048 shows no signs of the IVI.  However,  D2048r looks almost identical to A2048.  The $L_2$ norm of the difference of dye concentration fields between D2048r and D4096 almost exactly follows the norm of the difference between A2048 and D4096.

This shows that the IVI is a physical instability of this system.  It is not seen in the Dedalus simulations or the highest-resolution Athena simulation because the initial condition does not project sufficiently onto its unstable modes.  Errors in low resolution Athena simulations incorrectly excite perturbations unstable to the IVI.  Dedalus simulations, and the highest-resolution Athena simulation, suppress noise well enough the instability never becomes nonlinear.
In our phase-space diagram  (Figure~\ref{fig:schematic}), the lower-resolution Athena simulations do not properly follow the black line, and instead meander to the right, becoming unstable to the IVI.  D2048r is initialized to the right of IC', so it develops the IVI just like A2048.

As a final test, we started a Dedalus simulation from the output of an Athena simulation at $t=0$.  This tests whether dynamical evolution causes the IVI, rather than small differences between the implementation of the initial conditions. Although this introduced root mean squared differences in the horizontal velocity of $\approx 4\times 10^{-4}$ at $t=0$, the Dedalus simulation did not develop the IVI.

\subsubsection{Chaos}\label{sec:chaos}

At around $t\approx4$, D2048, D3072, and A16384 start to diverge exponentially from D4096 (Figure~\ref{fig:diff_1e5_2}).  The differences increase with a growth rate of about 2-3, much lower than the growth rate of 8 of the IVI found in the lower-resolution Athena simulations.  We interpret the differences between the simulations as due to chaos.  The faster divergence discussed in section~\ref{sec:IVI} is inconsistent with chaos since it is resolution dependent and only seen in low-resolution Athena simulations.

A system is chaotic if small differences between initial conditions grow exponentially in time.  To confirm the system is chaotic, we calculate a ``local-in-time'' Lyapunov exponent (i.e., growth rate).  We pick a time and simulation, and look for linearly unstable  perturbations.  This requires solving an eigenvalue problem. The largest unstable eigenvalue is the Lyapunov exponent.  Appendix~\ref{sec:eigenvalue} details this procedure.

This calculation does not inculde base-state time evolution (i.e. we consider a ``local-in-time'' calculation).  The most unstable eigenvector at a time $t_0$ might differ significantly from the most unstable eigenvector at a nearby time $t_0+\Delta t$.  Then it would be impossible for perturbations to grow at the Lyapunov exponent over times $\sim \Delta t$.  We interpret our ``local-in-time'' Lyapunov exponents as an upper bound on the growth rate of perturbations due to chaos (up to logarithmic corrections), and as a heuristic measure of the strength of chaos in this problem.

We calculated the Lyapunov exponent for D2048 with ${\rm Re}=10^5$ and $\Delta\rho/\rho_0=1$ at two times, $t=2.5$ and $t=4.5$.  We find Lyapunov exponents of $\lambda_{t=2.5}\approx 2.1$, and $\lambda_{t=4.5}\approx 3.7$.  Thus, the exponential growth of differences between either D2048, D3072, or A16384 and D4096 is consistent with chaos.  However, the growth rate of the differences between the lower-resolution Athena simulations and D4096 is much larger than the Lyapunov exponent.  These differences are inconsistent with chaos, instead being due to the IVI (section~\ref{sec:IVI}).

The simulations with $\Delta\rho/\rho_0=0$ do not appear to diverge from one another in the same way.  The highest-resolution Dedalus simulations converge at late times.  We also calculate the Lyapunov exponent for D1024 with ${\rm Re}=10^6$ and $\Delta\rho/\rho_0=0$ at $t=6$.  We find $\lambda_{t=6}\approx 0.4$.  Although this seems inconsistent with our finding that the Dedalus simulations approach each other with time, recall that this ``local-in-time'' calculation gives an upper bound on the growth rate due to chaos (up to logarithmic corrections).  Because the turnover time is 1, a Lyapunov exponent less than 1 suggests that small perturbations cannot grow before the background state changes substantially.  To show definitively that the $\Delta\rho/\rho_0=0$ solution is not chaotic, one should maximize the amplification of an initial perturbation over several turnover times, e.g., between $t=6$ and $t=9$ (for instance, using the adjoint method, e.g., \citealt{kerswell14}).

\subsubsection{Initial condition}\label{sec:perturb}

Although our chosen initial condition does not lead to the IVI for converged simulations, one might wonder if other initial conditions do lead to this instability.  We performed several Dedalus simulations that add low-amplitude white noise to the initial condition (e.g., see section~\ref{sec:IVI}).  None of these simulations develop the IVI.

We now consider a simulation in which we include perturbations to the initial condition with order unity amplitude and large wavelengths.
Equations~\ref{eqn:ICs} still hold for all quantities except the vertical velocity, which we now take to be
\begin{align}\label{eqn:perturb vz}
u_z &= A \left(\sin(2\pi{}x)+f(x)\right) \times \left[\exp\left(-\frac{(z-z_1)^2}{\sigma^2}\right) + \exp\left(-\frac{(z-z_2)^2}{\sigma^2}\right)\right],
\end{align}
where $f(x)$ includes Fourier modes two--ten.  Each mode receives a random phase and random amplitude uniformly distributed between -0.05 and 0.05.  Thus, $f(x)$ represents about a 10\% perturbation to the single sine mode initial condition.

\begin{figure}
\includegraphics[width=\columnwidth]{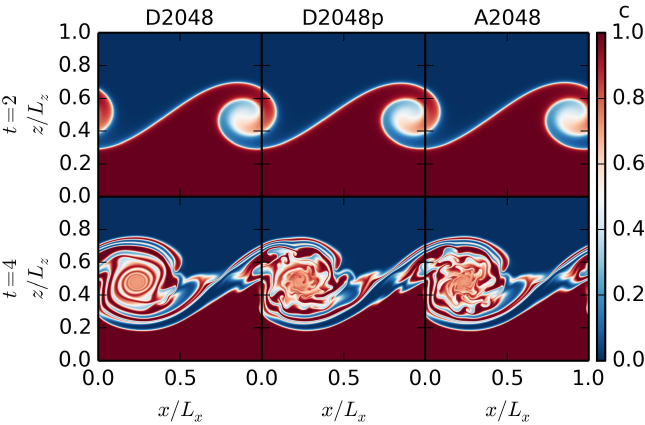}
  \caption{Snapshots of dye concentration field for ${\rm Re}=10^5$ and $\Delta\rho/\rho_0=1$.  D2048p is a Dedalus simulation with an initial vertical velocity that includes power over a range of Fourier modes (equation~\ref{eqn:perturb vz}), in contrast to the single mode initial conditions focused on throughout the rest of this paper.  At $t=2$ all solutions look the same, indicating that longest wavelength mode has the largest growth rate.  At $t=4$, D2048p has developed the IVI, as well as other deviations from the Dedalus \& Athena simulations away from the vortex.}\label{fig:IC}
\end{figure}

Figure~\ref{fig:IC} shows snapshots of the dye concentration field for this simulation, denoted D2048p, along with D2048 and A2048 for comparison.  At $t=2$, all three simulations look identical.  This indicates that the lowest wavenumber Fourier mode grows faster than the other modes included in our initial condition.

By $t=4$ the perturbations from the other Fourier modes produce significant changes to the dye concentration field.  D2048p now displays the IVI.  In addition, large differences appear away from the vortex, where the Dedalus and Athena simulations look almost identical.  Because the new initial condition does not respect the shift-and-reflect symmetry of the problem, the two half domains have different features (we only show the bottom half).

\subsubsection{Simulations without explicit diffusion}\label{sec:not explicit}

\begin{figure}
\includegraphics[width=\columnwidth]{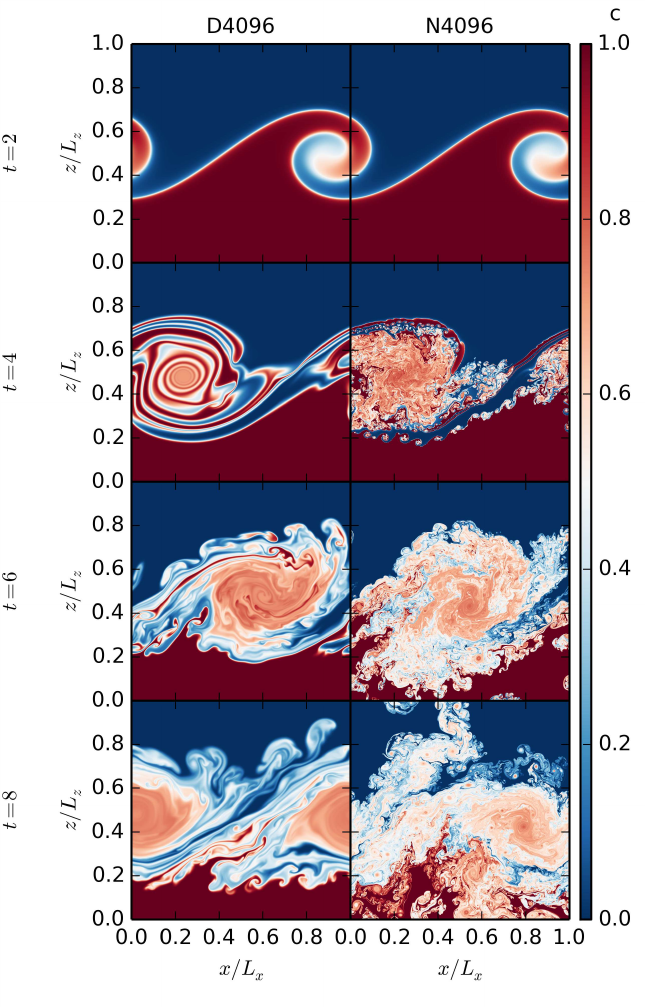}
  \caption{Snapshots of dye concentration field for $\Delta\rho/\rho_0=1$.  N4096 is an Athena simulation with no explicit diffusion.  For comparison, we also plot D4096 (${\rm Re}=10^5$).  Secondary instabilities occur very early at many locations in N4096.  By $t=6$, the simulation has broken its initial symmetry (we only plot the bottom half).  The secondary instabilities produce significant mixing, leading to greater entropy generation than in simulations with explicit diffusion (Figure~\ref{fig:entropy_1e5_2_nodiff}).}\label{fig:nodiff}
\end{figure}

Lastly, Figure~\ref{fig:nodiff} compares the resolved simulations at ${\rm Re}=10^5$ with an Athena simulation with horizontal resolution 4096 without explicit diffusion (N4096).  The simulation without explicit diffusion exhibits many secondary instabilities early in the evolution (between $t=2$ and $t=4$).  Unlike the lower-resolution simulations at ${\rm Re}=10^5$, the secondary instability is not limited to the IVI.  Instead, instabilities grow throughout the domain at locations of strong shear.

\begin{figure}
 \includegraphics[width=\columnwidth]{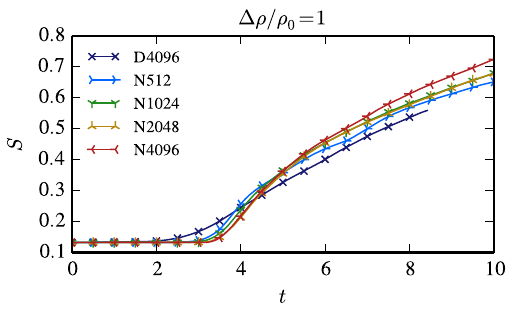}
  \caption{Volume-integrated dye entropy (equation~\ref{eqn:entropy}) as a function of time for simulations with $\Delta\rho/\rho_0 = 1$.  D4096 is run at ${\rm Re}=10^5$, and all simulations labeled with N are run with Athena with no explicit diffusion.  At early times, the highest-resolution runs without explicit diffusion have the lowest entropy.  However, at around $t=5$, the lower-resolution runs without explicit diffusion have lower entropy.  D4096 has the lowest entropy at late times.  This indicates that simulations without explicit diffusion have {\it greater} numerical mixing compared to simulations with explicit diffusion.  This becomes more prominent as the resolution increases.  By contrast, in the simulations without an initial density jump, explicit diffusion leads to more mixing, and for simulations without explicit diffusion, increasing resolution decreases mixing (Figure~\ref{fig:entropy_conv}).}\label{fig:entropy_1e5_2_nodiff}
\end{figure}

These instabilities shred apart the vortex, leading to vigorous mixing.  Figure~\ref{fig:entropy_1e5_2_nodiff} compares the volume-integrated dye entropy of Athena simulations with no explicit diffusion at different resolutions with D4096. Simulations without explicit diffusion produce almost no entropy until $t\approx 3.5$. At this time, the secondary instabilities start to cause diffusion at the grid-scale.  This generates entropy more rapidly than the explicit diffusion of D4096 (or any of the other simulations with explicit diffusion).  For $t>5$, the entropy of the simulations without explicit diffusion is larger than the entropy of D4096.  Paradoxically, the entropy increases as the resolution increases.  Our expectation is that the entropy generation should decrease as ${\rm Re}$ increases.  However, we do not have any resolved simulations with higher ${\rm Re}$ for comparison, so we cannot present evidence that this additional mixing is spurious.  But this problem shows that introducing an explicit diffusion in Athena can {\it decrease} the diffusion in the simulation.

\section{Conclusion}\label{sec:conclusion}

This paper describes several converged, nonlinear solutions to the Kelvin-Helmholtz (KH) problem.  By using a smooth initial condition and explicit diffusion, we demonstrate that solutions remain virtually identical (for constant initial density) or very similar (for an initial density jump of one) with resolution above a certain threshold.  This permits a well-defined reference solution for this problem, against which errors can be accurately estimated.  We verify this using two codes, Dedalus and Athena, with very different numerical methods (pseudo-spectral and Godunov, respectively).  Previous KH test problems either did not use smooth initial conditions, or did not include explicit diffusion.  Absent these two choices, the KH problem cannot be quantitatively compared between codes because the solutions depend sensitively on grid-scale errors and do not converge with increasing resolution.

We first study simulations with a constant initial density (section~\ref{subsec:unstratified-results}).  We find converged solutions to this relatively easy problem with Reynolds numbers (${\rm Re}$) as high as $10^6$.  
The solution is characterized by the continual roll-up of the initial vortex sheet, producing alternating filaments of unmixed material (Figure~\ref{fig:unstratified-dye}).  We find third-order convergence in both Dedalus \& Athena for simulations with ${\rm Re}=10^5$ (Figure~\ref{fig:error_1e5_1}), and better than second-order convergence in both codes for simulations with ${\rm Re}=10^6$ (Figure~\ref{fig:error_1e6_1}).

To quantify mixing in the simulations, we calculate the volume-integrated dye entropy as a function of time for several Reynolds numbers, as well as for Athena simulations without explicit diffusion (Figure~\ref{fig:entropy_conv}).  As the Reynolds number increases, the entropy generation decreases monotonically.  Similarly, as the resolution of Athena simulations without explicit diffusion increases, the entropy generation also decreases monotonically.  The entropy of one Athena simulation without explicit diffusion is very close to the entropy of the ${\rm Re}=10^6$ simulation, although the solutions show minor differences (Figure~\ref{fig:zoomin}).  These small differences indicate that the numerical diffusion in Athena does not act precisely as a physical diffusion from viscosity and/or thermal conductivity.  For certain applications however, assigning an effective Reynolds number to ideal fluid simulations may suffice.  This does not appear to be the case for KH simulations with density jumps, as we now discuss.

Including an initial density gradient aligned with the velocity gradient makes the problem much richer (section~\ref{sec:drat 2}).  The rolled-up vortex-sheet filaments becomes unstable in at least two ways: the inner vortex instability, and/or the outer filament instability (Figures~\ref{fig:stratified-dye} \& \ref{fig:schematic}).  The Dedalus simulations and highest-resolution Athena simulation only exhibit the outer filament instability, whereas the lower-resolution Athena simulations also exhibit the inner vortex instability.  Adding small amplitude noise to the initial condition does not produce the inner vortex instability in Dedalus, demonstrating that our chosen initial condition is not susceptible to this instability; instead, numerical errors seed the inner vortex instability throughout the evolution of the Athena simulations.  It is not surprising that Dedalus is more accurate than Athena for this smooth flow---the Godunov method is designed for simulating flows with shocks.  However, it is not well appreciated that the pseudo-spectral method is able to solve the full Navier-Stokes equations with Mach number order unity.

We use the $L_2$ norm to quantify the difference between dye concentration fields of different simulations, and find the inner vortex instability grows at a rate of $\approx 8$, independent of resolution (Figure~\ref{fig:diff_1e5_2}).  Furthermore, a Dedalus simulation initialized with an Athena state in the linear phase of the inner vortex instability develops the instability in the same way as Athena (Figure~\ref{fig:IVI}), demonstrating the physical, rather than numerical, nature of the instability. 

Adding a large ($\sim 10\%$ by energy) perturbation with multiple Fourier modes to the initial velocity in Dedalus can seed the inner vortex instability (section~\ref{sec:perturb}).  Although this suggests that the inner vortex instability is possibly generic for KH instabilities in astrophysics, we believe the single-mode initial condition discussed throughout the rest of this paper is still particularly valuable for a test problem. Because small numerical errors can produce large differences in the solution, one can assess by eye the fidelity with which a code is solving the fluid equations.  This KH test problem is difficult, which we believe makes it interesting.  In contrast, an unresolved KH problem is not a good test of fluid codes, because noise due to numerical errors can masquerade as higher-fidelity solutions.

The Dedalus simulations and highest-resolution Athena simulation also diverge from each other exponentially at late times, but with a much smaller growth rate $\approx 2-3$.  In section~\ref{sec:chaos} we calculate the maximum Lyapunov exponent of the flow, and argue that chaos drives the  divergence. The Lyapunov exponent represents the maximum possible rate of divergence of solutions due to chaos (up to logarithmic corrections).  At late times when the Dedalus simulations and highest-resolution Athena simulation begin to diverge, the Lyapunov exponent is $\approx 3.7$, so the divergence we see is consistent with chaos.  Because the system is chaotic, our solutions are not as accurate as the solutions with constant initial density.  We still find power-law convergence in the Dedalus simulations at fixed time (Figure~\ref{fig:diff_1e5_2}).  However, the amount of time that a solution maintains a fixed level of accuracy increases only logarithmically with resolution.

For the initial condition with a density jump, we also compare a high-resolution Athena simulation without explicit diffusion to our converged (within the limits of chaos) simulations with ${\rm Re}=10^5$.  Secondary instabilities pervade the simulation without explicit diffusion (Figure~\ref{fig:nodiff}).  The secondary instabilities cause enhanced mixing, and at late times, the simulations without explicit diffusion have higher entropy than the ${\rm Re}=10^5$ simulation (Figure~\ref{fig:entropy_1e5_2}).  Introducing explicit diffusion into Athena can {\it reduce} the diffusion in the simulation.  For this reason, we hypothesize (but cannot prove) that this small-scale structure is likely unphysical, and would not develop for any reasonable initial condition or Reynolds number.  This highlights that a solution with more small-scale structure is not necessarily better.

Although we only describe simulations with an initial density ratio of one, we have experimented with larger initial density ratios (e.g., 4). Preliminary investigation suggests that vigorous secondary instabilities become increasingly prominent as the density ratio increases, greatly enhancing mixing.  Though it's a common practice to leave out explicit dissipation to model the high Reynolds numbers relevant in astrophysics, our results suggest that including explicit diffusion may provide a very effective way to reduce diffusion in astrophysical simulations with very large density ratios.  We stress that these large density ratios are common in astrophysical problems such as star formation or galaxy formation.  Our results demonstrate just how subtle and computationally challenging it is to correctly capture mixing in these environments  (even restricting ourselves to hydrodynamics, which is likely a poor approximation).

There are many remaining questions left unanswered in this paper.  It is unclear how the Athena algorithm seeds the inner vortex instability.  We did not search for the critical perturbation amplitude that will cause a Dedalus simulation to exhibit the inner vortex instability.  Because of limited computer time, we did not find converged Dedalus or Athena simulations with $\Delta \rho/\rho_{0} = 1$ and ${\rm Re}=10^6$.  Perhaps, contrary to expectation, increasing the Reynolds number of the system does increase the entropy production, as found in the Athena simulations without explicit diffusion.  Future work should also test the Galilean invariance of these simulations, test initial conditions with an interface at an angle to the grid, and extend this analysis to larger density ratios.

We hope this study provides a well-posed test problem for future codes used in astrophysics.  It would be valuable to carry out this test problem with unstructured/meshless methods \citep[e.g.,][]{Springel2010,Duffell11,Hopkins2015} to understand their convergence properties on this challenging problem.  Toward this goal, we include the reference solutions to these KH problems in the supplementary material accompanying this paper.  Introducing smooth initial conditions and explicit diffusion allows us to calculate a converged reference solution and compare between codes.  The competing secondary instabilities for initial conditions with a density jump of one provides a stringent test of the fidelity with which a code solves the Navier-Stokes equations, making it a great test problem.

\section*{Acknowledgments}

\noindent{}The authors would like to thank Ramesh Narayan, Phil Hopkins, Paul Duffell, and C\'edric Beaume for helpful discussions.  D.L. is supported by the Hertz Foundation.  M.M. was supported by the National Science Foundation grant AST-1312651.  E.Q. is supported in part by a Simons Investigator Award from the Simons Foundation.  G.M.V acknowledges support from the Australian Research Council, project number DE140101960. J.S.O. is supported by a Provost's Research Fellowship from Farmingdale State College. The authors acknowledge the Texas Advanced Computing Center (TACC) at The University of Texas at Austin for providing HPC resources that have contributed to the research results reported within this paper.
This work used the Extreme Science and Engineering Discovery Environment (XSEDE allocations TG-AST140039, TG-AST140047, and TG-AST140083), which is supported by National Science Foundation grant number ACI-1053575.  Resources supporting this work were provided by the NASA High-End Computing (HEC) Program through the NASA Advanced Supercomputing (NAS) Division at Ames Research Center and the NASA Center for Climate Simulation (NCCS) at Goddard Space Flight Center.  This project was supported by NASA under
TCAN grant number NNX14AB53G.

\appendix

\section{Interpolation to a Common Grid}\label{sec:interpolate}

The grid points used in Dedalus and Athena differ slightly.  For a periodic simulation between $0$ and $L$ with spacing $\Delta x$, the Dedalus grid points are $\{ 0, \Delta x, 2\Delta x,\ldots,L-\Delta x\}$, whereas the Athena grid points are $\{\Delta x/2, 3\Delta x/s, \ldots, L-\Delta x/2\}$.  We use two spectrally accurate methods for interpolating Dedalus and Athena data to a common grid.  In several cases we test both methods and find excellent agreement.

Our first method is spectral interpolation.  The Dedalus data can be viewed as either $N=L/\Delta x$ values on grid points or $N$ Fourier coefficients.  We can pad the Fourier coefficients with zeros and transform to a grid of any uniform spacing. Going from $N$ to $2N$ points, we can compare every other entry to the Athena data.  In a second method, we multiply the Fourier coefficients by $\exp(i k_x\Delta x/2)$.  A Fourier transform then shifts the grid points by $\Delta x/2$, to align the Dedalus grid with the Athena grid. We follow the same procedure in the $z$ direction. 

Throughout this paper, we treat the Athena data (including initial conditions) as cell-centered data.  However, the data are actually volume-averaged.  The lowest-order differences between cell-centered and volume-averaged quantities scales as $\sim \Delta x^2$.  Thus, any errors associated with these differences should decrease with order 2.  In all cases studied here (i.e., Figures~\ref{fig:error_1e5_1}, \ref{fig:error_1e6_1}, \& \ref{fig:diff_1e5_2}), we find better-than second-order convergence.  This suggests that differences due to interpreting data as cell-centered rather than volume-averaged is not the dominant source of error.

\section{Convergence to a ``True'' Solution}\label{sec:convergence}

This paper describes a series of calculations of the nonlinear evolution of the KH instability, as a function of resolution and $\text{Re}$.  Without an analytic solution, we must assess the quality of the solutions carefully.  We make two assumptions to help interpret our results.

\begin{enumerate}
\item Dedalus and Athena converge to the same solution at fixed $\text{Re}$ as the resolution increases.  We refer to this unattainable ``Platonic ideal'' solution as the true solution.
\item The distance (given a choice of norm) between two solutions at different resolutions (for the same code), is larger than the distance between the higher-resolution simulation and true solution.
\end{enumerate}

Our simulations support these assumptions, but it is very difficult, if not impossible, to prove these statements.  The existence and uniqueness of solutions to the Navier-Stokes equations remains an active field of research \citep{fefferman2000}.

\begin{figure}
  \includegraphics[width=\columnwidth]{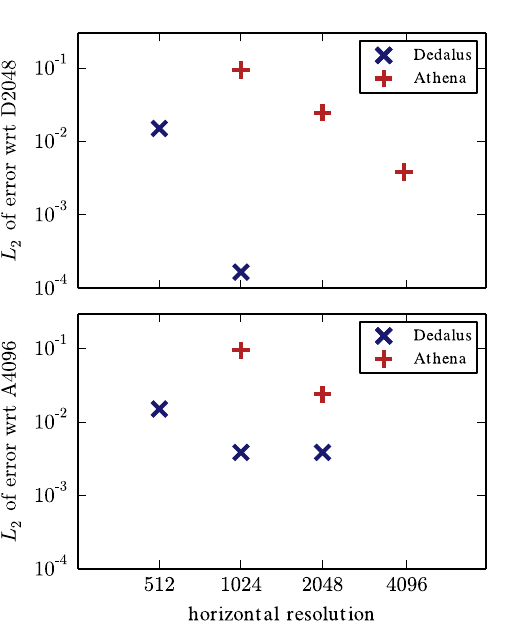}
  \caption{Differences between different solutions for ${\rm Re}=10^6$ and $\Delta\rho/\rho_0=0$ at $t=6$.  In the top panel, the Dedalus simulation with horizontal resolution 2048 (D2048) is assumed  to be the true solution, and in the bottom panel the Athena simulation with horizontal resolution 4096 (A4096) is assumed to be the true solution.  The error with respect to the assumed true solution is the $L_2$ norm of the difference of the dye concentration fields (equation~\ref{eqn:error}).  The top panel shows that both Athena and Dedalus are converging to the high-resolution Dedalus solution, supporting the assumption that it is close to the true solution.  In the bottom panel, the Athena solutions are converging to the high-resolution Athena simulation, but the Dedalus solutions are not.  This suggests that the Athena solution is further from the true solution than the Dedalus solutions.}\label{fig:distances}
\end{figure}

To support these assumptions, Figure~\ref{fig:distances} plots the relative differences between simulations with ${\rm Re}=10^6$ and $\Delta\rho/\rho_0=0$ at $t=6$ (described further in section~\ref{sec:1e6}).   The top panel assumes our highest-resolution Dedalus solution is the true solution. The bottom panel assumes our highest-resolution Athena solution is the true solution.  To assess the deviations, we plot the $L_2$ norm of the difference of dye concentration fields.  This allows us to define an error (alternatively a distance) between two solutions X and Y as
\begin{align}\label{eqn:error}
e({\rm X},{\rm Y}) = L_2(c_X-c_Y),
\end{align}
where $c_{\rm X}$ and $c_{\rm Y}$ are the dye concentration fields of solutions X and Y, respectively.  Figure~\ref{fig:distances} remains mostly unchanged if we compare lower Reynolds number simulations with ${\rm Re}=10^5$ and $\Delta\rho/\rho_0=0$, although the picture is more complicated for $\Delta\rho/\rho_0=1$ due to chaos (see section~\ref{sec:drat 2}).

Both the Athena simulations and the lower-resolution Dedalus simulations are converging to D2048.
The top panel of Figure~\ref{fig:distances} therefore suggests a true solution lives very close to D2048 (assumption (i)). The Athena simulations converge slower than the Dedalus simulations because in Dedalus spatial errors decrease exponentially.

The bottom panel of Figure~\ref{fig:distances} shows that A4096 is a worse approximation to the true solution.  This is because the Dedalus simulations are not converging to A4096.

One could argue that perhaps the Athena simulations are converging to a solution near A4096 and the Dedalus simulations are converging to a different solution near D2048.  However, this would require the error of the Athena simulations with respect to D2048 to stay constant as the resolution increases, contrary to the top panel.  Thus, we believe that both codes are converging to a true solution close to  D2048 (assumption (i)).

Presumably if Athena were run at very high resolutions, it would become closer to the true solution than D2048.  In this case, we hypothesize that both Dedalus and Athena simulations would converge to this very high-resolution Athena simulation.  For the range of resolutions examined in this paper, our highest-resolution Dedalus simulation is always closest to the true solution.

The main idea behind assumption (ii) is the convergence properties of the algorithms used in Athena and Dedalus.  Specifically, both codes are better than first-order accurate.  Imagine we somehow know the true solution to our problem, T.  If we run a high-resolution simulation, S1, we calculate the error,
\begin{align}
e({\rm S1},{\rm T}) \equiv E_1.
\end{align}
Now suppose we run another simulation S2 at double resolution.  If S1 and S2 are converging to T, then 
\begin{align}\label{eqn:accuracy}
e({\rm S2},{\rm T}) \equiv E_2 < \frac{E_1}{2},
\end{align}
where better-than first-order accuracy implies the inequality.  Athena is between second- and third-order accurate, so we expect $E_1/4 \le E_2 \le E_1/8$ for Athena.  Dedalus is exponentially accurate in space, and third-order accurate in time.  Thus, for Dedalus, we should expect $E_2 \le E_1/8 $.  Nevertheless, equation~\ref{eqn:accuracy} implies, via the triangle inequality,
\begin{align}\label{eqn:assumption 2}
e({\rm S1},{\rm S2}) > \frac{E_1}{2} > e({\rm S2},{\rm T}),
\end{align}
which shows that assumption (ii) holds.  One can check visually that equations~\ref{eqn:accuracy} \& \ref{eqn:assumption 2} hold for the simulations described in Figure~\ref{fig:distances}, assuming that T is very close to D2048.

\section{Lyapunov exponent calculation}\label{sec:eigenvalue}

One can write the equations of motion (equations~\ref{eqn:equations of motion}) as
\begin{align}\label{eqn:simplified}
\partial_t U = F(U),
\end{align}
where $U = (\rho,\vec{u},E)$ is the state vector.  Then infinitesimal perturbations to $U$ evolve according to the equation
\begin{align}
\partial_t \delta U = \left.\frac{\delta F}{\delta U}\right|_{U} \delta U,
\end{align}
where $\delta F/\delta U$ is the Fr\'{e}chet derivative, evaluated at $U(t)$.

To calculate the ``local-in-time'' Lyapunov exponent, we fix the state vector to its value at a specific time $t=t_0$.  The maximum Lyapunov exponent is the greatest eigenvalue of $(\delta F/\delta U)_{U(t_0)}$.  It is impractical to solve this eigenvalue problem directly---a 2D problem with resolution greater than 1000 in each direction generates very large matrices.  Instead, we solve an initial value problem by picking $\delta U(\tau=0)$, and evolving
\begin{align}\label{eqn:deltaU}
\partial_{\tau} \delta U = \left.\frac{\delta F}{\delta U}\right|_{U(t_0)}\delta U,
\end{align}
where $\tau$ should not be thought of as time, as we have fixed the background state $U$ to $t=t_0$.  The maximal Lyapunov exponent is
\begin{align}
\lambda = \lim_{\tau\rightarrow\infty} \log\left(\frac{||\delta U(\tau)||}{||\delta U(0)||}\right),
\end{align}
for some norm $||\cdot||$.  We choose $\sqrt{||\vec{u}||^2}$.  This is equivalent to the power method.

We solve equation~\ref{eqn:deltaU} in Dedalus using two methods.  Both methods give very similar Lyapunov exponents.  In the first, we directly evolve the linearized  equations~\ref{eqn:deltaU}.  We  treat terms independent of $U_0$ implicitly, and treat all other terms explicitly.  The second method uses an iteration. On each iteration, we evolve the full equations of motion (equation~\ref{eqn:simplified}) for $U_0+\delta U_i$ for a time $\Delta t\ll t_0$ to get a state we call $\tilde{U}_i(t_0+\Delta t)$.  The initial perturbation for the next iteration becomes $\delta U_{i+1} \propto \tilde{U}_i(t_0+\Delta t) - \Delta U_0$, but with norm $10^{-8}$.  $\Delta U_0 = U(t_0+\Delta t) - U(t_0)$ is the change in the unperturbed solution $U_0$ over the time $\Delta t$.  We normalize after each iteration to ensure the perturbations stay linear.

In both cases, we initialize the calculation with a guess of random noise.  After substantial evolution, the system begins to execute limit cycles (or seems to be close to a limit cycle for $\Delta\rho/\rho_0=1$ at $t=2.5$).  We report the growth averaged over a limit cycle.

\bibliographystyle{mn2e}
\bibliography{kh}

\label{lastpage}
\end{document}